\documentclass[%
 reprint,
 amsmath,amssymb,
 aps,
prb,
]{revtex4-2}

\setcitestyle{super}
\usepackage{graphicx}
\usepackage{dcolumn}
\usepackage{bm}

\def\<{\langle}
\def\>{\rangle}

\begin{document}

\preprint{DHpaper}

\title{Anomalous electrodynamics and quantum geometry in the \\ Dirac-Harper model for a graphene bilayer}

\author{Abigail Timmel}
\author{E. J. Mele}
\email{mele@physics.upenn.edu}
\affiliation{
 Department of Physics and Astronomy\\
 University of Pennsylvania \\
 Philadelphia, PA 19104
}

\date{\today}

\begin{abstract}
	Graphene bilayers with layer antisymmetric strains are studied using the Dirac-Harper model for a pair of single layer Dirac Hamiltonians coupled by a one-dimensional moir\'e-periodic interlayer tunneling amplitude. This model hosts low energy, nearly dispersionless bands near charge neutrality that support anomalous polarizations of its charge multipole distributions. These are analyzed introducing a generalized Berry curvature that encodes the field-induced dynamics of multipole fields allowed in a chiral medium with time reversal symmetry. The formulation identifies a reciprocity relation between responses to layer-symmetric and layer-antisymmetric in-plane electric fields and reveals momentum-space quantum oscillations produced by a spatial pattern of band inversions on the moir\'e scale.
\end{abstract}

\maketitle

\section{Introduction}

The electrodynamic properties of electrons in crystals can include anomalous responses due to the momentum dependence of internal spin and orbital attributes in Bloch bands. There has been interest in interpreting these as geometric phenomena associated with a quantum metric.\cite{Resta, Neupert} This concept generalizes to driving fields that vary in space.  When a smooth spatial variation lowers the symmetry of the system, it can activate responses that would be forbidden for spatially uniform fields.  These spatially dispersive responses are sometimes expressed as responses of a system to the field {\it gradients}. Artificial structures fabricated by stacking atomically thin layers are of great current interest in this regard, and in these systems the field gradient is replaced by a discretized set of driving fields on separate layers.\cite{Morell,Stauber,Son}  In twisted graphene bilayers discretized coupled layer responses can greatly exceed spatially dispersive responses found in conventional chiral materials.\cite{Park}

Here we formulate coupled-layer electrodynamic responses in Bloch bands as quantum geometric quantities.  We find that charge multipole distributions in the bilayer have dynamics that are naturally represented by a generalization of the Berry curvature to a {\it class} of curvature forms \cite{Provost} in a parameter space which combines position and momentum.  This approach generalizes the concept of an anomalous charge velocity induced by Berry curvature to a wider class of multipole electrodynamic responses. We illustrate this for the specific case of a graphene bilayer with a one-dimensional moir\'e periodicity induced by layer-antisymmetric strain.

The paper is organized as follows.  In Section II for completeness we briefly review the construction of the Dirac-Harper model for our model system, a shear-strained one-dimensional moir\'e structure.\cite{DH}  In Section III we describe features of the low energy spectrum for this model which reveals a set of weakly-dispersing modes with charge peaks that counterpropogate in the moir\'e as a function of the moir\'e-transverse crystal momentum.  In Section IV we characterize multipole densities which associate this counterpropagation with a net  polarization of a multipole distribution.  In Section V we formulate a generalized Berry curvature that describes the anomalous response of these multipole densities to the electric field.  In Section VI we examine the symmetry constraints of both the multipole susceptibilities (Section V) and the static multipole polarizations (Section IV) which turn out to be distinct phenomena.  In Section VII we explain the origin of oscillatory behavior found numerically in calculations of the multipole susceptibilities.

\section{Dirac-Harper Model}

\begin{figure}[!h]
	\begin{center}
		\includegraphics[width=\columnwidth]{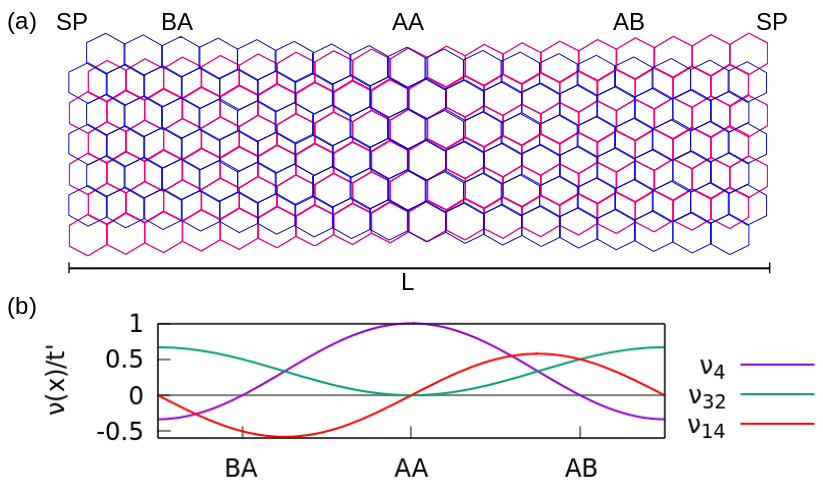}
	\end{center}
	\caption{(a) A section of the shear-strained (AC) moir\'e structure studied in this paper, which is taken to repeat indefinitely in the $x$ and $y$ directions.  Typical moir\'e lengths are around 900 lattice constants in the $x$ direction.  (b)  Scalar amplitudes of the three mass terms $\gamma_{4}$, $\gamma_{14}$, and $\gamma_{32}$ normalized by the interlayer coupling constant $t'$.}
	\label{structure}
\end{figure}

The system we study is a strained graphene bilayer in which the stacking is modulated in a single crystallographic direction $x$ (Figure~\ref{structure}a) while retaining short range lattice periodicity in the orthogonal coordinate $y$. Here for completeness we summarize the formulation of this problem as a spinorial Harper model.\cite{DH}

The Hamiltonian for this system includes a spatially varying term $\boldsymbol{\nu}(x)$ which couples the two layers and a set of kinetic terms ${\bf t}$ which couple adjacent cells in the same layer.  The conserved crystal momentum $k_y$ allows this system to be separated into a family of one-dimensional problems along the $x$ coordinate where the effects of the $y$-directed couplings are absorbed into a $k_y$ dependence of the kinetic energy terms.  The result is a generalization the scalar Harper equation to a spinorial Dirac-Harper problem \cite{DH}
\begin{equation}
	\label{harper}
	{\bf t}(k) \boldsymbol{\psi}_{n+1} + {\bf t}^\dagger (k) \boldsymbol{\psi}_{n-1} + {\bf t}_0\boldsymbol{\psi}_n + \boldsymbol{\nu}(x)\boldsymbol{\psi}_n = \epsilon \boldsymbol{\psi}_n \\
\end{equation}
Here ${\boldsymbol{\psi}}_n$ are four component fields with amplitudes on two sublattices ($a$ and $b$)  and two layers ($1$ and $2$) with the ordering  ${\boldsymbol{\psi}}^T = (a_1,b_1,a_2,b_2)$.  The integers $n$ in Equation~\ref{harper} index  the primitive cells of the microscopic graphene lattice along the $x$ direction.  The ${\bf t}(k)$ are block diagonal $4 \times 4$  matrices that can be derived from a nearest neighbor tight-binding kinetic energy for the individual graphene layers.\cite{supplement}  Defining the Pauli matrices $\sigma_i$ and $\tau_i$ acting on sublattice index and layer index respectively, the products $\sigma_x\tau_0 = \gamma_{15}$ and $\sigma_y\tau_0 = \gamma_{25}$ form a basis for ${\bf t}$.

The interlayer coupling $\boldsymbol{\nu}(x) = \boldsymbol{\nu}(x + L) $ is a smooth spatial variation through three high-symmetry stacking configurations: $AA$ where equivalent sublattices in the two layers align vertically, $AB$ where $a_1$ sublattices eclipse $b_2$ sublattices, and $BA$ where $b_1$ sublattices eclipse $a_2$ sublattices.  The smoothest interpolation between these three configurations is found using a Fourier series in the lowest one dimensional $G$-vectors
$$ \boldsymbol{\nu}(x) = \sum_G \boldsymbol{\nu}_G e^{iGx} $$
with matrix-valued coefficients $\boldsymbol{\nu}_G$.  The high-symmetry stacking configurations utilize three $\gamma$ matrices: $\gamma_4 = \sigma_0\tau_x$, $\gamma_{14} = \sigma_y\tau_y$, and $\gamma_{32} = \sigma_x\tau_x$, so the series can be re-expressed as
$$ \boldsymbol{\nu}(x) = \sum_\alpha \nu_\alpha(x) \gamma_\alpha $$
for scalar $\nu_\alpha(x)$, where $\gamma_\alpha \in \left\{\gamma_4,\gamma_{14},\gamma_{32}\right\}$.  The amplitudes $\nu_\alpha(x)$ are plotted in Figure~\ref{structure}b.

The symmetries of this model depend on the orientation of the modulation direction with respect to the underlying graphene lattice, and this crucial feature is encoded in the exact form of the kinetic terms ${\bf t}$. An important limiting case studied here occurs for modulation along the armchair (AC) direction, which is achieved by imposing a layer-antisymmetric shear strain (Figure~\ref{structure}a).  This produces a chiral structure with the two-fold rotational symmetries around three orthogonal axes, $C_{2x}$, $C_{2y}$, and $C_{2z}$, but no mirror symmetries.  In contrast, modulation along the zigzag (ZZ) direction requires uniaxial strain and the resulting structure respects mirror symmetries.  The presence of these mirror symmetries is a nongeneric feature of the DH model which occurs only for special orientations of the modulation direction.  The kinetic terms are expressed in a low-energy effective theory by taking $\boldsymbol{\psi}_{n\pm 1} \approx \boldsymbol{\psi}_{n}$ in Equation~\ref{harper} and linearizing the combination ${\bf t}(k) + {\bf t}^\dagger(k) + {\bf t}_0$ in small momentum $q$ about a Dirac point.  For AC, this linearization gives a kinetic energy operator $\hbar v_F(\gamma_{15}q_x - \gamma_{25}q_y)$ where $v_F = \sqrt{3}at/2$, $t$ is the intralayer coupling constant, and $a$ is the graphene lattice constant.  To compare, ZZ is a $\pi/2$ rotation of AC which swaps the roles of $\gamma_{15}$ and $\gamma_{25}$.

\section{Low Energy Spectra}

\begin{figure}[!h]
	\begin{center}
		\includegraphics[width=\columnwidth]{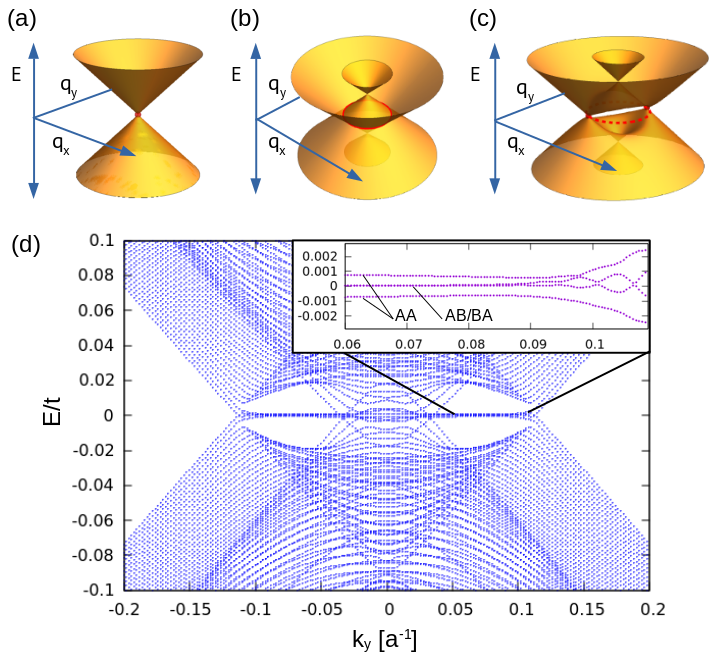}
	\end{center}
	\caption{(a) Dirac cones produced by the linearized kinetic energy.  (b) Dirac cones split by $\gamma_4$ to intersect on a critical ring.  (c) Gapping of the critical ring by $\gamma_{14}$ except at two points in the $q_y$ axis.  (d) Full moir\'e spectrum for the AC structure. {\it Inset:} Magnification of the low-energy bands within the gap.  Each band is two-fold degenerate, and there is a total of four $AA$ peaked states and four $AB/BA$-peaked states.}
	\label{spectrum}
\end{figure}

Figure~\ref{spectrum}d shows the band structure of the AC structure as a function of $k_y$.  Features of the spectrum can be understood in the long moir\'e limit directly from the commutation relations of the kinetic terms with the mass terms $\boldsymbol{\nu}(x)$ introduced by the interlayer tunneling.\cite{DH}  In the absence of interlayer coupling, the spectra consist of a pair of layer-degenerate Dirac cones shown in Figure \ref{spectrum}a.  In a semi-classical picture, modification of this spectrum by local mass terms describes behavior of Bloch electrons localized to specific regions within the moir\'e.  The $\gamma_4$ mass term describes same-sublattice hopping between the two layers, characteristic of an $AA$-registered region.  This mass term commutes with both kinetic terms, splitting two hybridized Dirac cones in energy so that they intersect on a ring in momentum space (Figure \ref{spectrum}b) which we refer to as the critical ring.  The mass term $\gamma_{14}$ describes an asymmetry in the coupling between opposite sublattices on the two layers which occurs at a generic position in the superlattice cell and is strongest in the $AB$ and $BA$ regions.  This term anticommutes with $\gamma_{15}$, gapping out the critical ring everywhere a except at two nodal points where the critical ring intersects the $q_y$ axis (Figure \ref{spectrum}c).

We can use these local four-band features to analyze the full moir\'e spectrum by noticing that in the long-moir\'e limit, the slowly-varying stacking allows modes to localize in regions of nearly constant stacking.  Referring to Figure~\ref{structure}b, the $AA$ region is a location where two out of three mass terms have extrema.  Indeed, we can readily identify the width of the region supporting zero-energy states in the full moir\'e spectrum (Figure \ref{spectrum}d) as the peak width of the critical ring in $AA$.  This critical ring occurs at $|q| = \hbar v_F/t' a \equiv q_c$ where $t'$ is the interlayer coupling strength.  The gap opened around zero energy with two point closures at $k_y = \pm q_c$ arises from $\gamma_{14}$ changing its sign in this region.

\begin{figure}[!h]
	\begin{center}
		\includegraphics[width=\columnwidth]{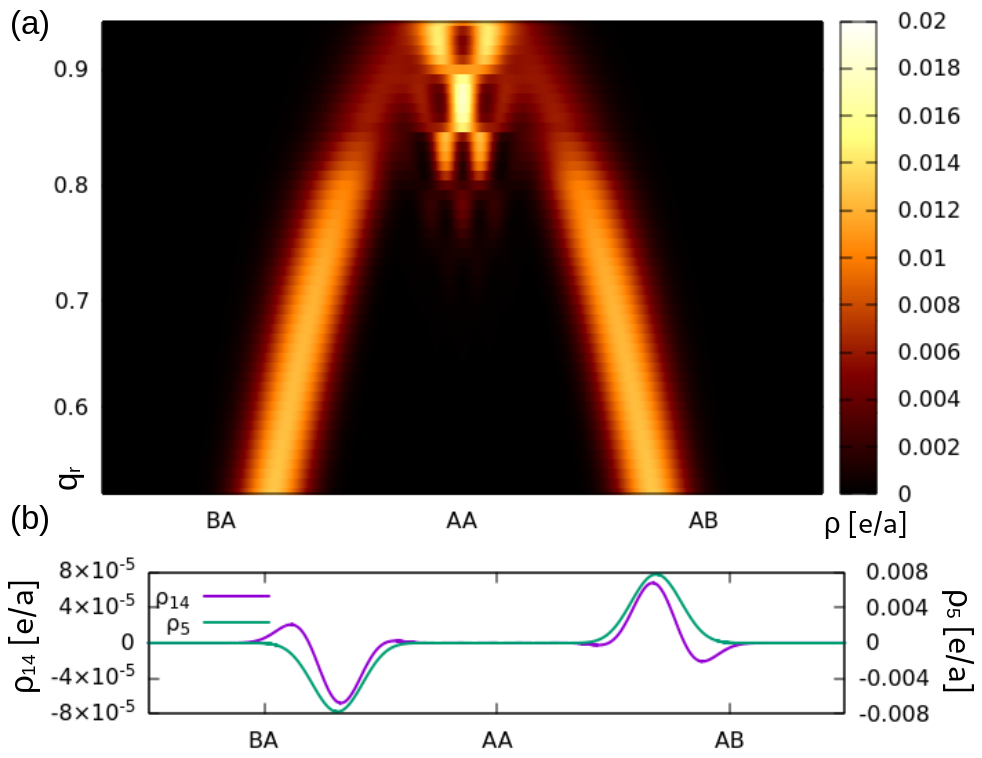}
	\end{center}
	\caption{(a) Anomalous drift of the charge peaks in $AB/BA$-peaked bands as a function of $q_r \equiv k_y t' a/\hbar v_F$.  The gap closure occurs at $q_r = 1$. (b) Multipole densities $\rho_{14}$ and $\rho_5$ at $q_r = 0.52$ tagging the peaks with opposite sign.}
	\label{densities}
\end{figure}

As shown in the inset of Figure~\ref{spectrum}d, a set of nearly dispersionless states appears within this gap.  These modes can be understood as produced by sign changes of the interlayer couplings as a function of lateral position $x$.\cite{DH}  One set of modes associated with a sign change of the $\gamma_{14}$ mass confines charge density to the $AA$ region of the moir\'e.  A second set of modes arises from a sign change of the $\gamma_4$ mass term giving a $k_y$-dependent potential that produces a charge density (Figure~\ref{densities}a) with two peaks that counterpropagate as a function of $k_y$.   These features migrate from the $AB$ and $BA$ regions towards the $AA$ region as $k_y$ approaches the gap closure at $k_y/q_c \equiv q_r = 1$.  In a semiclassical description, this counterpropagation could be driven by an electric field $\hbar \dot k_y  = -e E_y$. This counterpropagation transports no charge along $x$ but {\it can} be associated with transport of higher order multipole distribution.

%

\section{Multipole polarization}

We begin by inspecting multipole distributions that are antisymmetric in $x$ because these quantities tag the counterpropogating peaks (Figure~\ref{densities}a) with opposite signs and associate them with net polarization of a charge-neutral multipole field.  The lowest nontrivial multipole distribution that respects $C_{2x}$, $C_{2y}$, and $C_{2z}$ symmetries of the shear-strained AC moir\'e is an octopole.  A multipole distribution that is odd in $x$ can therefore be produced by a quantity that has the local rotational symmetry of a quadrupole in the $yz$ plane.  Such an octopolar distribution is already explicitly present in the DH model; the mass term $\hat \gamma_{14} \, \sin(G x)$ is a product of a quadrupolar operator $\hat \gamma_{14}$ with an antisymmetric function of $x$ that distinguishes $AB$ and $BA$ registered regions.  The multipole distribution $\rho_{14}(x) = \boldsymbol{\psi}^\dagger(x)\gamma_{14}\boldsymbol{\psi}(x)$ manifests this property in an observable by coupling the quadrupolar $\gamma_{14}$ to an antisymmetric spatial distribution (Figure~\ref{densities}b).  Similarly, a quadrupole density locally measured by the sublattice- and layer- odd operator $\hat \gamma_{5} = \sigma_z  \tau_z$ describes a local layer- and sublattice-polarization that is also spatially antisymmetric (Figure~\ref{densities}b).  We refer to transverse polarizations intrinsic to these antisymmetric multipole distributions as anomalous polarizations.

\section{Electrodynamic responses}
The response of an anomalous multipole polarization to an applied electric field can be expressed as a quantum geometric quantity. In a Bloch band with broken time reversal or inversion symmetry, an anomalous velocity for charge is produced by the Berry curvature $\<\partial_{\lambda_\mu}u|\partial_{\lambda_\nu} u\> - c.c.$.  To extend this to multipole dynamics, we examine the generalizated curvature tensor
\begin{equation}
	\omega_m = \< \tfrac{\partial u}{\partial \lambda_\mu}|M|\tfrac{\partial u}{\partial \lambda_\nu}\> - \< \tfrac{\partial u}{\partial \lambda_\nu}|M|\tfrac{\partial u}{\partial \lambda_\mu}\>
	\label{omega_m}
\end{equation}
where $M \equiv m_{\alpha\beta}\otimes\mathbb{I}_N$ and $m_{\alpha\beta}$ is a local multipole operator weighting the sublattice and layer degrees of freedom.  For $m_{\alpha\beta} = \mathbb{I}$, we recover the usual Berry curvature responsible for the anomalous Hall conductance. More generally, $\omega_m$ can describe the anomalous dynamics of a charge-neutral multipole distribution.  A mathematical discussion of this curvature is given in the Supplementary Information.\cite{supplement}

The parameters $\lambda$ specify a tangent space for a state in the Bloch band which will determine the association of this curvature with a physical response function.  To identify appropriate choices for these parameters for the Dirac-Harper model, we observe that the one-dimensional moir\'e structure is most naturally described using its two extended coordinates: a momentum coordinate in the $y$ direction and a {\it spatial} coordinate in the $x$ direction.  The two dual coordinates $k_x$ and $y$ are insignificant: zone-folding reduces the width the Brillouin zone $G_x$ by a factor of the inverse moir\'e length $1/L$, and the $y$ coordinate does not appear in the DH equation.  For definiteness we choose $k_x= 0$, promoting the model with symmetries that can be only weakly violated at order $1/L$.

We decompose the extended spatial coordinate by introducing three parameters, each of which controls the origin of its respective mass term in the Hamiltonian.  These spatial parameters $x_\alpha$ appear in $\boldsymbol{\nu}(x)$ as
\begin{equation*}
	\boldsymbol{\nu}(x)= \sum_{\alpha} \nu_{\alpha}(x+x_\alpha) \gamma_{\alpha}
\end{equation*}
This parameterization allows us to vary the origin for each mass field individually.  Each $x_\alpha$ paired with $k_y$ parameterizes a two-dimensional subspace of the projective Hilbert space.  Note that the ground state lies in the one-dimensional sector at $x_\alpha = 0$ where the mass fields are pinned by the moir\'e structure. Nonetheless the two-dimensional {\it tangent space} generated by differential changes $\partial_{x_\alpha}$, $\partial_{k_y}$ will encode an observable property of these states which we denote
\begin{equation}
	K_\alpha^{(M)} = \< \tfrac{\partial u}{\partial k_y}|M|\tfrac{\partial u}{\partial x_\alpha}\> - c.c.
	\label{curvature}
\end{equation}
Notice that $K_\alpha^{(M)}$ must have the symmetry of an octopole to respect $C_2$ symmetry.  This is satisfied if $m$ has the symmetry of a $z$-directed dipole, and we use $\tau_z$ as a simple and useful example. 

\begin{figure}[!h]
	\begin{center}
		\includegraphics[width=\columnwidth]{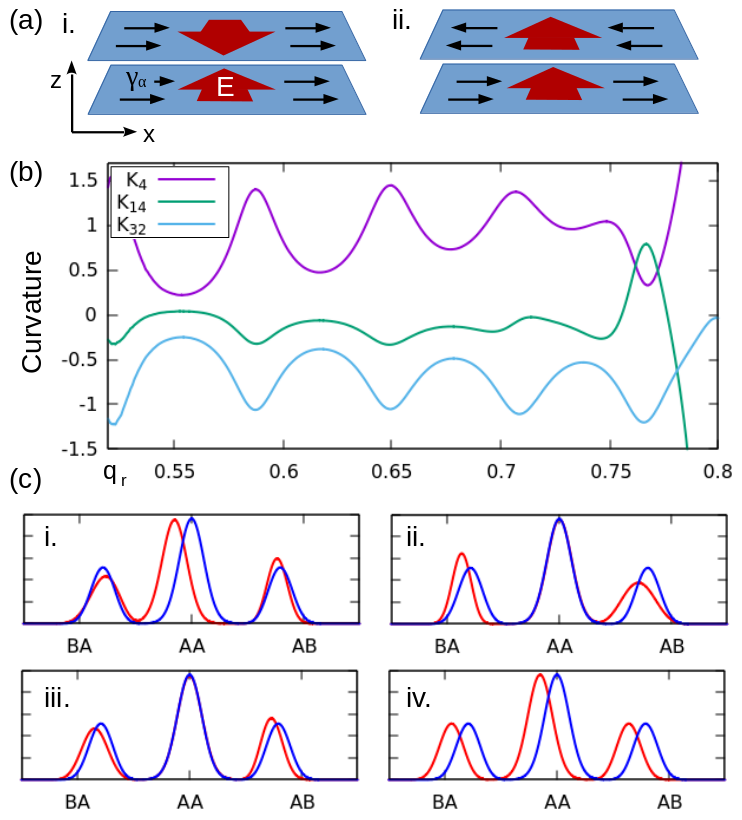}
	\end{center}
	\caption{(a) Reciprocal systems described by i. Equation~\ref{antisym_E} and ii. Equation~\ref{antisym_response}. (b) Curvatures $K^{(\tau_z)}_\alpha$ for each of the mass channels.  (c) Change of the charge density from i. $x_{14}$, primarily affecting the $AA$ charge peak, ii. $x_{4}$, primarily affecting the $AB/BA$ peaks, iii. $x_{32}$, causing little displacement, and iv. all the mass terms together, inducing a constant shift. }
	\label{curvature_plots}
\end{figure}

The triad of operators $\partial_{x_\alpha},\partial_{k_y}, \tau_z$ can now be identified with two different response functions.  The operator $\partial_{k_y}$ couples the band to a perturbing electric field applied in the $y$-direction, and the operator $\partial_{x_\alpha}$ arises from the observable $\hbar\dot{k}_x^{(\alpha)} = [\partial_{x_\alpha},H]$ which describes an $x$-directed force on a component of the wavefunction coupled to the $\alpha$-th mass term.  The coupling to individual mass terms will be explained in more detail in the next paragraph, but for now it is adequate to think of this as a force acting on component of the charge density.  To introduce the third member of the triad, $\tau_z$, we can consider either a layer-antisymmetric electric field or a layer-antisymmetric response.  For the former, $\tau_z$ enters in the driving term $E \tau_z \hat{y}$, and the force is given by
\begin{equation}
	f_{y\alpha} = i eE \left( \sum_{m\neq n}\frac{\<n|[\frac{\partial}{\partial x_{\alpha}},H]|m\>\<m|\tau_z|\frac{\partial n}{\partial k_y}\>}{\epsilon_m-\epsilon_n} - c.c. \right) 
	\label{antisym_E}
\end{equation}
where the first subscript labels the force by an induced lattice effect, to be described later.  For the latter, $\tau_z$ weights the observable $i\partial_{x_\alpha}$ with opposite sign in each layer:
\begin{equation}
	f_{x\alpha} = i eE \left( \sum_{m\neq n}\frac{\<n|[\tau_z\frac{\partial}{\partial x_\alpha},H]|m\>\<m|\frac{\partial n}{\partial k_y}\>}{\epsilon_m-\epsilon_n} - c.c. \right) 
	\label{antisym_response}
\end{equation}
Both of these can be formally expressed as manifestation of  {\it same}  generalized curvature $K_\alpha^{(\tau_z)}$ in Equation~\ref{curvature}, and thus they are equivalent.\cite{supplement}  Numerical calculation for each mass channel $\alpha$ is plotted as a function of $q_r$ in Figure~\ref{curvature_plots}b.

Equations \ref{curvature}, \ref{antisym_E} and \ref{antisym_response} are the main results of this work.  These results can be regarded as variants of the constitutive relations that describe natural optical activity in chiral media.\cite{Nonguez2, Nonguez3, Morell, Opticsbook}  Natural optical activity arises from a correlated electric dipole-quadrupole response driven by an optical field, while the multipole susceptibilities obtained here describe an analogous coupling of two different charge multipoles in the static limit.  Each mass channel describes a force acting on the component of the charge density confined by the associated mass term.  This is demonstrated by examining the differential changes of the charge density produced by shifts of each $x_\alpha$, shown explicitly in Figure~\ref{curvature_plots}c.  The AA peaks defined by the $\gamma_{14}$ mass inversion are shifted by $x_{14}$ (i) and the $AB/BA$ peaks defined by $\gamma_4$ inversion are shifted with $x_{4}$ (ii).  Coupling to  $x_{32}$ produces almost no effect (iii) because $\gamma_{32}$ does not confine charge in the moir\'e cell via a sign change.\cite{DH}  The combination of all three $x_\alpha$ produce uniform charge shift (iv), so the sum of $K_\alpha$ over mass channels a gives the anomalous susceptibility for polarization of a multipole-weighted density ($\tau_z$ in this example).

The equivalence of Equations~\ref{antisym_E} and \ref{antisym_response} identifies a reciprocity relation stating that a force induced by a multipole-weighted field is equal in magnitude to the multipole-weighted force induced by a uniform field.  For our example of a $z$-directed dipole $\tau_z$, these responses are driven by a layer-antisymmetric field $\vec{E}_\leftrightarrows$ and a layer symmetric field $\vec{E}_\rightrightarrows$ respectively.  The layer-antisymmetric field can be thought of the low-frequency limit of a plane wave polarized in the $y$ direction.  For case of $\vec{E}_\leftrightarrows$ (Equation~\ref{antisym_E}), the response summed over mass channels produces a force that acts to depin charge in the negative $x$ direction as shown in Figure~\ref{curvature_plots}c iv.  The $E_\rightrightarrows$ case (Equation~\ref{antisym_response}) corresponds to a similar displacement of charge, but weighted in opposite directions in the two layers.

Displacements of various components of the charge density can produce a counterforce acting on the lattice to restore the moir\'e to an equilibrium stacking arrangement with respect to the charge.  The coupling of lattice strain fields to the applied field must respect the $C_2$ symmetries, so we can deduce the allowed combinations.  Focusing our discussion on uniform layer-antisymmetric displacements, the field $\vec{E}_{\leftrightarrows}$ (odd in $y$ and $z$) couples to displacement along $y$ (also odd in $y$ in $z$), and the field $\vec{E}_{\rightrightarrows}$ (odd only in $y$) couples to displacement along $x$ (odd in $x$ and $z$).  The subscripts in Equations~\ref{antisym_E} and \ref{antisym_response} are now clear: $f_y$ ($f_x$) induces a layer shear along $y$ ($x$).  Intuitively, the $y$-directed shear transports the full moir\'e pattern along the $x$ direction\cite{Xiao} to bring the lattice back in phase with the charge, and the $x$-directed shear counters the layer antisymmetric distortion of the charge along $x$ induced by the initial electrodynamic response.  We can write a general $y$-directed electric field $E_y(z)$ as a sum of $E_\leftrightarrows + E_\rightrightarrows$, where $E_\leftrightarrows = [E_y(+z)-E_y(-z)]/2$ and $E_\rightrightarrows = [E_y(+z)+E_y(-z)]/2$.  Then the induced layer shear acts at a general angle $\phi$ in the $xy$ plane, which using reciprocity of the curvature is given by $\phi = \tan^{-1}(f_y/f_x) = \tan^{-1}(E_\leftrightarrows/E_\rightrightarrows)$.  $\phi=0$ corresponds to opposing layer displacements in the $x$ direction driven by a uniform electric field in the $y$ direction.

\section{Symmetry Requirements}

The response functions described by $K_\alpha$ are symmetric under time-reversal $\mathcal{T}$: since $K_\alpha$ is imaginary and contains only one momentum coordinate, it accumulates two cancelling sign changes under $\mathcal{T}$.  This means that  the anomalous response described by $K_\alpha$ does not require $\mathcal{T}$-breaking to be present.  In contrast to the usual Berry curvature which vanishes everywhere in a Bloch band that respects both time-reversal and inversion symmetry, the generalized susceptibility studied here vanishes everywhere for systems with time-reversal and {\it mirror} symmetry.  We can gain additional insight using $C_{2x}\mathcal{T}$ symmetry, which is a local symmetry in the parameter space $(x,k_y)$.  The symmetry of $m_{\alpha\beta}$ under this operation determines whether the curvature $K^{(M)}$ or its associated multipole density $\rho_m$ must vanish locally in the mixed momentum-position space.  For the curvature, one finds \cite{supplement}
\begin{equation*}
	\< \tfrac{\partial u}{\partial k_y}|M|\tfrac{\partial u}{\partial x_\alpha}\>
	 = \< \tfrac{\partial u}{\partial k_y}|M'|\tfrac{\partial u}{\partial x_\alpha}\>^*
\end{equation*}
for $M' = (C_{2x}^\dagger M C_{2x})^*$.  When $M' = M$, the curvature is equal to its complex conjugate and therefore it vanishes everywhere.  Indeed this is the case for the charge susceptibility $M=\mathbb{I}$, but the generalized extensions can incorporate multipoles such as $\tau_z$ where $M' = -M$, allowing this response to be nonzero.  For the multipole density $\rho_m$, the same manipulation gives $ \rho_m = \rho_{m'}^* $ where $m'$ is the $x$-projected version of $M'$.  Since $\rho_m$ is real, this quantity vanishes when $m' = -m$.  Thus we find that the symmetry of $m$ under $C_{2x}\mathcal{T}$ divides multipole densities into two distinct classes: those that can have nonvanishing density but vanishing curvature and those that have vanishing density but a nonvanishing curvature.  The anomalous response measured by Equation~\ref{curvature} therefore is physically distinct from the construction obtained by antisymmetrically-tagging the counterpropagating peaks shown in Figure~\ref{densities}a. Dynamical effects occur in quantities with no bulk accumulation of a multipole density, and quantities with bulk accumulation can only display a static polarization.

For unitary multipole operators,\cite{supplement} the generalized Berry curvature is related to the standard Berry curvature by diffeomorphisms on the Hilbert space.  This is a consequence of the Darboux theorem which states that any symplectic two-form can be re-expressed in a standard form on an open set via a coordinate change.  Thus, we arrive at an alternative interpretation of the generalized Berry curvature as the standard Berry curvature computed on a different sector of the Hilbert space.  As an example, we can put the $\tau_z$ curvature into standard form by complex-conjugation of the second-layer amplitudes of the wavefunctions, which has the effect of negating the second-layer component of the complex curvature.  The new wavefunctions do not have obvious physical interpretation, but the standard form of the Berry curvature lends computational and conceptual power.  Calculations of the generalized Berry curvature can be very difficult in non-standard form, but after applying the diffeomorphism, a standard Wilson loop can be utilized.\cite{supplement}

\section{Quantum oscillations}

\begin{figure}
	\medskip
	\begin{center}
		\includegraphics[width=\columnwidth]{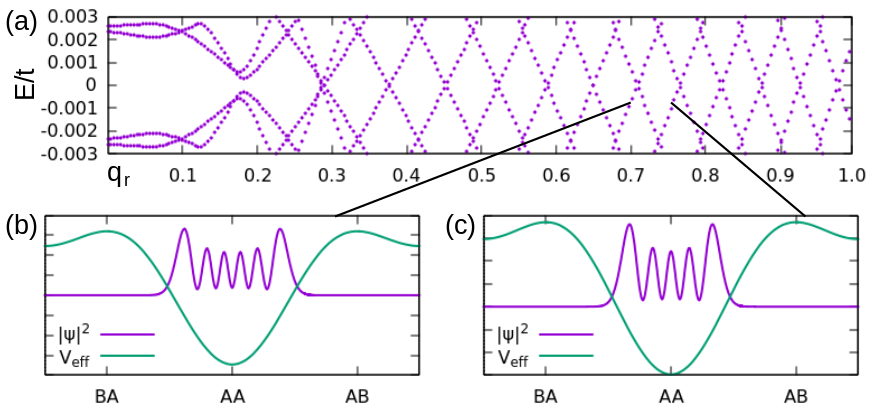}
	\end{center}
	\caption{(a) Zero energy crossings of the tight-binding spectrum for the $\nu_4(x)$ potential.  Tight-binding wavefunctions plotted against the effective potential for (b) $k_y=.08$ and (c) $k_y=.086$.  As wavefunctions from consecutive bands, they differ by one peak.}
	\label{veff}
\end{figure}

All three of the curvatures shown in Figure~\ref{curvature_plots}b oscillate as a function of $k_y$.  This behavior manifests oscillations in the energies of the $AB/BA$ bands, partially shown in Figure~\ref{spectrum}d, which are encoded in the response function via the velocity terms in the Kubo formula.  The observed oscillations have a smoothly evolving period which is nearly constant over the width of the band-inverted region $|q_r|<1$ in momentum space.  We can understand this in a minimal model by examining the low-energy behavior of spatially varying $\gamma_4$ potential, shown in Figure~\ref{veff}.

One might be tempted to interpret the spacing of zero energy crossings as a result of zone-folding of the critical ring (Figure~\ref{spectrum}b) into the first moir\'e Brillouin zone.  Such an effect arises from a constant $\gamma_4$ potential, which we can inspect numerically using a tight-binding model.  This calculation reveals that a non-negligible $\pm q_x$ anisotropy of the critical ring produces two independently-evolving series of zero-energy crossings in the band structure that rapidly converge near the edge of the critical ring $q_r=\pm 1$.\cite{supplement}  Such behavior is quite different from the slowly modulated period of the oscillations we observe in the $\tau_z$ curvature.  By contrast, a spatially {\it varying} $\gamma_4$ potential produces a nearly uniform pattern of band crossings (Figure~\ref{Veff}a) which closely matches the period of the curvature oscillations.  This behavior can be understood by squaring the low-energy Dirac equation,
\begin{equation*}
	\left(i\hbar v_F \gamma_{15} \tfrac{\partial}{\partial x} + \hbar v_F k_y \gamma_{25} + \nu_4(x)\gamma_4\right)\psi = \epsilon\psi
\end{equation*}
to form the effective potential problem
\begin{equation}
	\tfrac{\partial^2}{\partial x^2}\psi = \left(-\tfrac{\nu_4(x)^2}{\hbar^2 v_F^2} + k_y^2\right)\psi \equiv V_\text{eff}\,\psi
	\label{Veff}
\end{equation}
The spacing of zero-crossings in this situation arises from a resonance condition within a basin where $V_\text{eff}$ is negative.  The wavefunctions produced by the $\nu_4(x)\gamma_4$ potential in a tight-binding calculation (Figure~\ref{veff}b) oscillate in the region where $V_\text{eff}$ is negative and decay in regions where it is positive.  Wavelengths of the oscillating part of the wavefunctions match the value of $\sqrt{V_\text{eff}}$ averaged over each period.  Consecutive zero energy band crossings differ by a single wavefunction oscillation within the $V_\text{eff}$ basin, which is indicative that the spacing of the zero-energy crossings is a consequence of wavefunction resonance within this effective potential.

These observations about the effective potential produced by a pure $\nu_4$ potential carry over to the full DH Hamiltonian.  The mass term proportional to $\gamma_{14}\sin(Gx)$ has the effect of opening a gap around the lowest energy band manifold without changing the spacing of these zero energy crossings.\cite{supplement}  Away from the saddle point configuration in the stacking texture ($SP$), $\gamma_{32}$ decreases to zero and has only a perturbative effect.  Therefore, oscillatory behavior in the energies is inherited from the $k_y$ spacing of zero energy crossings intrinsic to the $\gamma_4$ part of the potential.   Although the full moir\'e $AB/BA$ wavefunctions decay within the positive effective potential region, the sensitivity of the energies to these special values of $k_y$ suggests that the wavefunctions retain a small oscillatory part stretching over the positive potential region flanked by the $AB/BA$ charge peaks.  Interestingly these oscillations would be absent from chiral models for twisted bilayer graphene which exclude this coupling to enforce a chiral symmetry.\cite{TV} Here we see that they control a physically measurable response, and measurement of these oscillations can be used to determine the scale of the breaking of the chiral symmetry.

\section{Conclusion}

The curvature forms developed here can also be applied to two dimensional twisted bilayers.  In this case the DH theory generalizes to a two dimensional theory in a moir\'e supercell and the forces driving the multipole distributions analogous to Equations \ref{antisym_E} and \ref{antisym_response} are similarly promoted to two dimensional vectors obtained by integrating over a population in the 2D folded Brillouin zone. The simplest case couples the driving fields to the total charge density recovering the charge pumping  phenomena proposed for twisted graphene bilayers.\cite{Fujimoto, Xiao} The generalized Berry curvature provides a unified formulation that associates this electromechanical response with the more general problem of manipulating multipole distributions using applied fields. The generalized constitutive relations for electric field-driven multipole responses also provide a geometric formulation of the natural optical activity of twisted graphene bilayers studied experimentally \cite{Park} and theoretically \cite{Morell,Son,Stauber} and to chiral Weyl semimetals.\cite{Kargarian}

This work was supported by the Department of Energy under grant DE-FG02-84ER45118.

\nocite{*}

\pagebreak
\widetext
\begin{center}
	\textbf{\large Supplementary Information: Anomalous Electrodynamics and Quantum Geometry in Graphene Heterobilayers}
\end{center}
\setcounter{equation}{0}
\setcounter{figure}{0}
\setcounter{table}{0}
\setcounter{section}{0}
\setcounter{page}{1}
\makeatletter
\renewcommand{\theequation}{S\arabic{equation}}
\renewcommand{\thefigure}{S\arabic{figure}}
\renewcommand{\thetable}{S\arabic{table}}
\renewcommand{\bibnumfmt}[1]{[S#1]}
\renewcommand{\citenumfont}[1]{S#1}

\section{Structures}

The kinetic terms ${\bf t}_0$ and ${\bf t}$ in main text derive from the tight-binding model for single-layer graphene
$$ {\bf t}_0 =
t\begin{pmatrix}
	0 & 1 & 0 & 0 \\
	1 & 0 & 0 & 0 \\
	0 & 0 & 0 & 1 \\
	0 & 0 & 1 & 0
\end{pmatrix}, \quad
{\bf t} =
t\begin{pmatrix}
	0 & e^{i(\frac{k_x}{2} + \frac{\sqrt{3}k_y}{2})} & 0 & 0 \\
	e^{i(\frac{k_x}{2} - \frac{\sqrt{3}k_y}{2})} & 0 & 0 & 0 \\
	0 & 0 & 0 & e^{i(\frac{k_x}{2} + \frac{\sqrt{3}k_y}{2})} \\
	0 & 0 & e^{i(\frac{k_x}{2} - \frac{\sqrt{3}k_y}{2})} & 0
\end{pmatrix}$$
The interlayer coupling $\boldsymbol{\nu}(x)$ is given by a series in the lowest $G$-vectors fixing the three high-symmetry stacking configurations $AA$, $AB$, and $BA$.  The result is
$$ \boldsymbol{\nu}(x) = t'\left(\tfrac{1}{3}(1+\cos(2\pi x/L))\gamma_4 + \tfrac{1}{3}(1-2\cos(2\pi x/L))\gamma_{32} + \tfrac{1}{\sqrt{3}}\sin(2\pi x/L)\gamma_{14}\right)$$

\section{Generalized Berry Curvature}

The usual Berry curvature on the projective Hilbert space $\mathbb{CP}^{4N}$ takes the form 
$$ \omega =  \sum_{i,\alpha} \mathrm{d} u_{i\alpha}^* \wedge \mathrm{d} u_{i\alpha}$$
where $u_{i\alpha}$ denotes the wavefunction component for the $\alpha$th sublattice of the $i$th cell.  We insert the multipole by contracting a tensor $m_{\alpha\beta}$ on the sublattice degrees of freedom
\begin{equation*}
	\omega_m = \sum_{i,\alpha,\beta} m_{\alpha\beta}\;\mathrm{d}u^*_{i\alpha}\wedge\mathrm{d}u_{i\beta}
\end{equation*}
When $\omega_m$ is applied to tangent vectors $\partial_{\lambda_\nu}$ and $\partial_{ \lambda_\mu}$ at $|u\>$ (where $\lambda_\mu$, $\lambda_\nu$ parametrize the band subspace) we obtain the familiar form of the Berry curvature weighted by the multipole $M \equiv m_{\alpha\beta}\otimes\mathbb{I}_N$ in Equation~\ref{omega_m} of the main text.

As long as $m_{\alpha\beta}$ is invertible, the generalized curvature is related to the usual Berry curvature by a diffeomorphism.  This can be appreciated by noting that the generalized Berry curvature is a symplectic form on $\mathbb{CP}^{4N}$, and by the Darboux Theorem for every neighborhood $U$ there is a choice of coordinates in which $\omega_m$ takes the standard form of the Berry curvature $\omega$.  The diffeomorphism generally changes the sector of the Hilbert space associated with a Bloch band, implying that integration is performed over different submanifolds of the Hilbert space and yields independent results.  Furthermore, unless the multipole operator satisfies certain conditions, the diffeomorphism may map to unnormalized wavefunctions.  Assuming that the multipole operator is Hermitian, the condition for preserving normalization is that the multipole operator must have eigenvalues of unit modulus, i.e. no scaling of coordinates occurs, only a unitary transformation.  When this condition is satisfied, we may use the diffeomorphism to obtain a set of wavefunctions on which the usual calculational techniques and topological analysis related to the Berry curvature can be applied to yield results for the multipole-weighted curvature.  In particular, a Wilson loop calculated over the new wavefunctions provides a simple method for calculating the generalized curvature.


To give a concrete example, when $m=\tau_z$, the coordinate change is complex conjugation of the second-layer amplitudes:
\begin{equation*}
	u'_{i\alpha} = \left\{\begin{array}{r}
		u_{i\alpha} \quad \alpha \in {a_1,b_1} \\
		u^*_{i\alpha} \quad \alpha  \in {a_2,b_2}
	\end{array}\right.
\end{equation*}
We see that
$$ \omega_{\tau_z} = \sum_{i,\alpha,\beta} (\tau_z)_{\alpha\beta}\;\mathrm{d}u^*_{i\alpha}\wedge\mathrm{d}u_{i\beta} = \sum_{i,\alpha} \mathrm{d} {u'}_{i\alpha}^* \wedge \mathrm{d} u'_{i\alpha}$$

We can thus view the $\omega_{\tau_z}$ curvature integrated over a Bloch band as equivalent to the standard Berry curvature integrated over a {\it different} sector of the Hilbert space, obtained by complex conjugating all second-layer amplitudes in the wavefunctions.

\section{Local Antiunitary Symmetry}

Here we show in detail the conditions for vanishing curvature and multipole densities under a local antiunitary symmetry $U\mathcal{K}$, where $\mathcal{K}$ denotes complex conjugation and $U$ is a unitary operator.  First, we choose an orthonormal basis of $U\mathcal{K}$ eigenstates with eigenvalues $+1$.  In the case of shear-strained graphene, the energy bands are two-fold degenerate, so we can construct these explicitly with the following algorithm.  Readers who are satisfied that such a basis can be constructed may skip this part.

\medskip

First we pick an arbitrary state $|v\>$ in the band subspace.  Since $U\mathcal{K}$ is a symmetry of the Hamiltonian, $U\mathcal{K}|v\>$ is an eigenstate with the same energy.  Then we construct within the band subspace
$$ |u_1\> = \tfrac{1}{\sqrt{2}}(|v\> + U\mathcal{K}|v\>), \quad |u_2\> = \tfrac{1}{\sqrt{2}}e^{i\pi/2}(|v\> - U\mathcal{K}|v\>) $$
These are eigenstates of $U\mathcal{K}$ both with eigenvalue $+1$.  To make them orthogonal, we observe that
$$ \<u_1|u_2\> = \tfrac{1}{2}e^{i\pi/2}(\<v|v\> - \<v^*|v^*\> + \<v^*|U^\dagger|v\> - \<v|U|v^*\>) $$
$$ = \tfrac{1}{2}e^{i\pi/2}(\<v^*|U^\dagger|v\> - \<v|U|v^*\>)  = e^{i\pi/2}\mathrm{Im}(\<v^*|U^\dagger|v\>)$$
This means that $|u_1\>$ and $|u_2\>$ are orthogonal if and only if $\<v^*|U^\dagger|v\>$ is real.  We can enforce this by replacing $|v\> \rightarrow e^{-i\theta/2}|v\>$ where $\theta = \mathrm{arg}(\<v^*|U^\dagger|v\>)$ in the above construction of the $|u\>$.

\medskip

In this basis $|u\>$, we now have that
$$ U\mathcal{K}|u\> = |u\> \;\;\; \Rightarrow \;\;\; |u^*\> = U^\dagger|u\> $$
The unitary part $U$ can then be inserted into the curvature form as follows
$$ \< \tfrac{\partial u}{\partial k_y}|M|\tfrac{\partial u}{\partial x_\alpha}\> = \< \tfrac{\partial u}{\partial k_y}|UU^\dagger M UU^\dagger|\tfrac{\partial u}{\partial x_\alpha}\> = \< \tfrac{\partial u^*}{\partial k_y}|U^\dagger M U|\tfrac{\partial u^*}{\partial x_\alpha}\>= (\< \tfrac{\partial u}{\partial k_y}|(U^\dagger M U)^*|\tfrac{\partial u}{\partial x_\alpha}\>)^* \equiv  (\< \tfrac{\partial u}{\partial k_y}|M'|\tfrac{\partial u}{\partial x_\alpha}\>)^*$$
which is the result from the main text.

For the multipole density, we follow a similar procedure on the $x$-projected wavefunctions $u(x)$.  Since we assume $U\mathcal{K}$ is local in $x$, we have
$$ U\mathcal{K} u(x) = u(x) \;\;\; \Rightarrow \;\;\; u^*(x) = U^\dagger u(x)$$
Then the multipole density satisfies
$$ \rho_m = u^*(x) m u(x) = u^*(x) UU^\dagger m UU^\dagger u(x) = u(x) U^\dagger m U u^*(x)$$
$$= (u^*(x)(U^\dagger m U)^* u(x))^* \equiv (u^*(x) m' u(x))^* = \rho^*_{m'}$$
as desired.

\section{Kubo Formulae}

Here we derive the curvature form (Equation~\ref{curvature} of main text) from Equations~\ref{antisym_E} and \ref{antisym_response} of the main text.  Starting with Equation~\ref{antisym_E}, we have
$$ f_{y\alpha} = i eE \left( \sum_{m\neq n}\frac{\<n|[\frac{\partial}{\partial x_{\alpha}},H]|m\>\<m|\tau_z|\frac{\partial n}{\partial k_y}\>}{\epsilon_m-\epsilon_n} - \frac{\<\frac{\partial n}{\partial k_y} |\tau_z |m\>\<m|[\frac{\partial}{\partial x_\alpha},H]|n\>}{\epsilon_m-\epsilon_n}\right) $$
$$ = i eE \left( \sum_{m\neq n}\frac{\<n|\left((\epsilon_m - \epsilon_n)\frac{\partial}{\partial x_{\alpha}} + \frac{\partial \epsilon_m}{\partial x_\alpha}\right)|m\>\<m|\tau_z|\frac{\partial n}{\partial k_y}\>}{\epsilon_m-\epsilon_n} - \frac{\<\frac{\partial n}{\partial k_y} |\tau_z |m\>\<m|\left((\epsilon_n - \epsilon_m)\frac{\partial}{\partial x_{\alpha}} + \frac{\partial \epsilon_n}{\partial x_\alpha}\right)|n\>}{\epsilon_m-\epsilon_n}\right) $$
Since $|n\>$ and $|m\>$ are orthogonal, the $\tfrac{\partial \epsilon_n}{\partial x_\alpha}$ terms drop out.  Then we can cancel the energy denominator, giving
$$ = i eE  \sum_{m\neq n}\left(\<n|\tfrac{\partial}{\partial x_{\alpha}}|m\>\<m|\tau_z|\tfrac{\partial n}{\partial k_y}\> + \<\tfrac{\partial n}{\partial k_y} |\tau_z |m\>\<m|\tfrac{\partial}{\partial x_{\alpha}}|n\>\right) $$
Integrating the first term by parts, we have
$$ = i eE  \sum_{m\neq n}\left(-\<\tfrac{\partial n}{\partial x_{\alpha}}|m\>\<m|\tau_z|\tfrac{\partial n}{\partial k_y}\> + \<\tfrac{\partial n}{\partial k_y} |\tau_z |m\>\<m|\tfrac{\partial n}{\partial x_{\alpha}}\>\right) $$
Now we can add and subtract $\<\tfrac{\partial n}{\partial x_{\alpha}}|n\>\<n|\tau_z|\tfrac{\partial n}{\partial k_y}\> = \<\tfrac{\partial n}{\partial k_y}|\tau_z|n\>\<n|\tfrac{\partial n}{\partial x_{\alpha}}\>$ to complete the sum.
$$ = i eE  \sum_{m}\left(-\<\tfrac{\partial n}{\partial x_{\alpha}}|m\>\<m|\tau_z|\tfrac{\partial n}{\partial k_y}\> + \<\tfrac{\partial n}{\partial k_y} |\tau_z |m\>\<m|\tfrac{\partial n}{\partial x_{\alpha}}\>\right)  = i eE  \left(-\<\tfrac{\partial n}{\partial x_{\alpha}}|\tau_z|\tfrac{\partial n}{\partial k_y}\> + \<\tfrac{\partial n}{\partial k_y} |\tau_z |\tfrac{\partial n}{\partial x_{\alpha}}\>\right) $$
as desired.  Equation~\ref{antisym_response} follows a similar derivation,
$$f_{x\alpha} = i eE \left( \sum_{m\neq n}\frac{\<n|[\tau_z\frac{\partial}{\partial x_\alpha},H]|m\>\<m|\frac{\partial n}{\partial k_y}\>}{\epsilon_m-\epsilon_n} - \frac{\<\frac{\partial n}{\partial k_y} |m\>\<m|[\tau_z \frac{\partial}{\partial x_\alpha},H]|n\>}{\epsilon_m-\epsilon_n}\right)$$
$$= i eE \left( \sum_{m\neq n}\frac{\<n|\left((\epsilon_m-\epsilon_n)\tau_z\frac{\partial}{\partial x_\alpha}+\tau_z\frac{\partial \epsilon_m}{\partial x_\alpha}\right)|m\>\<m|\frac{\partial n}{\partial k_y}\>}{\epsilon_m-\epsilon_n} - \frac{\<\frac{\partial n}{\partial k_y} |m\>\<m|\left((\epsilon_n-\epsilon_m)\tau_z\frac{\partial}{\partial x_\alpha}+\tau_z\frac{\partial \epsilon_n}{\partial x_\alpha}\right)|n\>}{\epsilon_m-\epsilon_n}\right)$$
$$= i eE \sum_{m\neq n}\left( \<n|\tau_z\tfrac{\partial}{\partial x_\alpha}|m\>\<m|\tfrac{\partial n}{\partial k_y}\> + \<\tfrac{\partial n}{\partial k_y} |m\>\<m|\tau_z\tfrac{\partial}{\partial x_\alpha}|n\>\right)$$
$$= i eE \left( -\<\tfrac{\partial n}{\partial x_\alpha}|\tau_z|\tfrac{\partial n}{\partial k_y}\> + \<\tfrac{\partial n}{\partial k_y} |\tau_z|\tfrac{\partial n}{\partial x_\alpha}\>\right)$$
again as desired.

\section{Energy Oscillations}

\begin{figure}[!h]
	\begin{center}
		\includegraphics[width=.6\columnwidth]{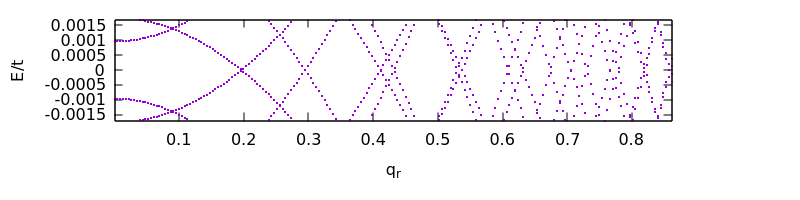}
	\end{center}
	\caption{Low energy bands for a zone-folded critical ring in a constant $\gamma_4$ potential, showing the two independently-evolving band spacings.}
	\label{const4}
\end{figure}

\begin{figure}[!h]
	\begin{center}
		\includegraphics[width=.6\columnwidth]{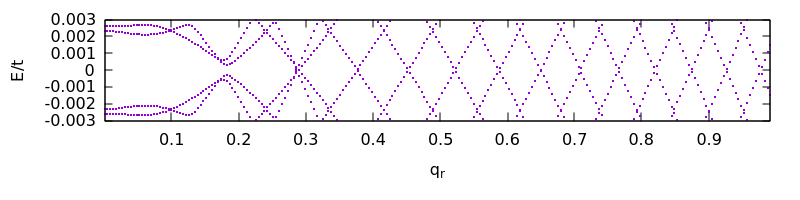}
	\end{center}
	\caption{Low energy bands for a varying $\gamma_4$ potential, displaying a single slowly-evolving band spacing in contrast with the constant $\gamma_4$ potential.}
	\label{vary4}
\end{figure}

\begin{figure}[!h]
	\begin{center}
		\includegraphics[width=.4\columnwidth]{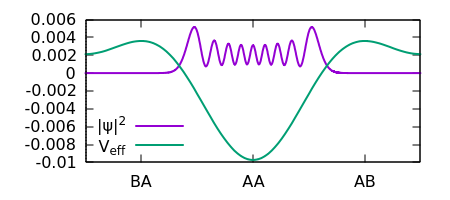}
		\includegraphics[width=.4\columnwidth]{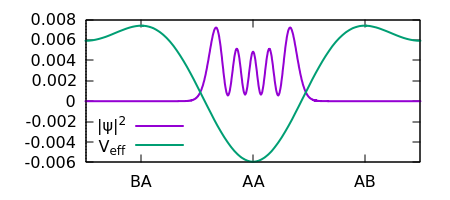}
	\end{center}
	\caption{Effective potentials and numerical charge densities for $k_y = 0.06$ and $k_y = .086$ indicating close agreement as the range of the $AA$ basin changes.}
	\label{S_veff}
\end{figure}

Here we offer further evidence that the oscillations found in the curvatures and energies originate from an effective potential constructed out of the $\gamma_{4}$ part of the potential.  To clarify the different character of the low-energy spectrum of a {\it constant} versus {\it varying} $\gamma_4$ potential, these are plotted in Figures~\ref{const4} and \ref{vary4} respectively.  One can observe that for the constant potential, not only do the spacings of the bands decrease rapidly to the critical ring edge, but there are two independently-evolving spacing periods.  This is because the nodal ring, formed by vertical displacement of the Dirac cones, is not perfectly symmetric.  Under zone-folding, bands displaced from $\pm G$ do not perfectly coincide but instead evolve out of phase.  In contrast, the {\it varying} $\gamma_4$ potential gives rise to a single, slowly evolving spacing period.  The qualitative difference suggests different mechanisms determining the energy spacing in these two situations.  The close agreement between the numerical wavefunctions and the effective potential formulated in Equation~\ref{Veff} of the main paper for differing values of $k_y$ (Figure~\ref{S_veff}) gives strong evidence that the spacing for the varying potential is determined by a resonance condition within the effective potential.

\begin{figure}[!h]
	\begin{center}
		\includegraphics[width=.6\columnwidth]{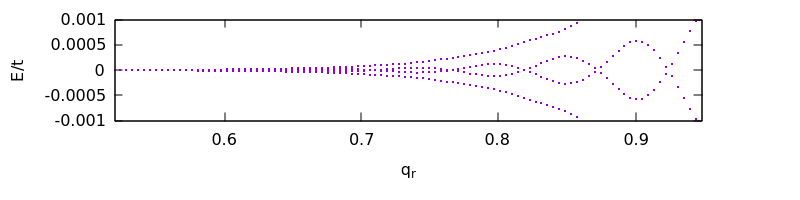}
	\end{center}
	\caption{Low energy bands with the $\gamma_4$ and $\gamma_{14}$ components of the moir\'e potential.  The zero-energy crossings line up with those of the varying $\gamma_4$ potential.}
	\label{vary4_sin14}
\end{figure}

The preservation of the $\gamma_4$ zero energy crossings under incorporation of the $\gamma_{14}$ part of the potential (Figure~\ref{vary4_sin14}) shows that this mechanism carries over to the full moir\'e.  Similary between this spectrum and the low bands of the full moir\'e shown in Figure 1 of the main paper support our claim that the $\gamma_{32}$ part of the potential only exerts perturbative influence on this part of the spectrum.

\nocite{*}

\end{document}